\documentclass{article}
\usepackage{jheppub}
\usepackage[dvipsnames]{xcolor}
\usepackage{amsmath,amssymb,amsfonts,slashed,cleveref}
\usepackage{fixmath} 
\usepackage{physics}

\usepackage{mathrsfs}  
\usepackage{graphicx}
\usepackage{appendix}

\usepackage{tcolorbox}
\tcbuselibrary{breakable}

\usepackage{tikz}
\usetikzlibrary{arrows.meta, positioning}

\newcommand{\fermiacc}{\texttt{FERMIACC}}

\definecolor{jsonbg}{RGB}{246,248,252}
\definecolor{jsonframe}{RGB}{120,140,180}
\definecolor{jsonkey}{RGB}{35,92,170}
\definecolor{jsonval}{RGB}{180,72,32}

\newcommand{\jk}[1]{\textcolor{jsonkey}{\texttt{#1}}}
\newcommand{\jv}[1]{\textcolor{jsonval}{\texttt{#1}}}

\newcommand{\jline}[2]{%
  \noindent\hspace*{#1}\parbox[t]{\dimexpr\linewidth-#1\relax}{#2}\par
}
\newcommand{\jsonblock}[1]{%
  \par\smallskip
  \noindent
  \setlength{\fboxsep}{8pt}%
  \fcolorbox{jsonframe}{jsonbg}{%
    \parbox{0.95\linewidth}{\ttfamily\small\raggedright\setlength{\parfillskip}{0pt} #1}%
  }%
  \par\smallskip
}

\begin{document}

\title{The \texttt{FERMIACC}: Agents for Particle Theory}

\author[a]{Prateek Agrawal,}
\author[a,b]{Nathaniel Craig,}
\author[b]{Amalia Madden,}
\author[a]{and I\~nigo Valenzuela Lombera}

\affiliation[a]{Department of Physics, University of California, Santa Barbara, CA 93106, USA}

\affiliation[b]{Kavli Institute for Theoretical Physics, University of California, Santa Barbara, CA 93106, USA}

\abstract{We present the \fermiacc, a scaffolded reasoning model built on OpenAI agents designed to
autonomously generate and quantitatively validate theory hypotheses for high energy physics data at scale. 
}

\maketitle


\section{\label{sec:intro} Introduction}

Artificial intelligence is poised to dramatically accelerate progress across the sciences. Within high energy physics, there is a colloquial narrative that this progress will take the form of an `AI Einstein', a reasoning large language model (LLM) capable of working out the theory of Nature through compute alone. While some of the problems of high energy physics may yield in this fashion (where existing data, thought experiments, and consistency conditions yield a unique answer), most will not. In the sixteen orders of magnitude between the weak scale and the Planck scale governed by effective field theory, decoupling guarantees that infinitely many distinct theories are nonetheless consistent with data at fixed precision and energy. The challenge here is not that we lack a theory compatible with all current data; it is that we have altogether too many.

In this respect, progress across high energy physics requires both making the most of the data at hand and acquiring new data at higher precision and energy. What is needed here is less an `AI Einstein' and more an `AI Fermi': a reasoning model that can characterize and navigate the space of viable theories, systematically testing and discriminating among them. Such a model would deploy optimal methods across fields and analyze experimental data in heretofore-unimaginable ways.
A central goal for such a reasoning model would be to autonomously generate and quantitatively validate plausible hypotheses to explain features in high energy physics data. Delivering this at scale would allow the exploration of innumerable hypotheses across innumerable experiments, maximizing the impact of existing datasets and dramatically accelerating the pace of scientific discovery. At the same time, such a reasoning model must overcome the limitations facing the current generation of LLMs, such as susceptibility to hallucinations and challenges in maintaining long-horizon reasoning continuity and verifiability.

 At the time of writing, LLMs have reached the point where this vision seems to be within reach. Frontier LLMs now regularly achieve high-level performance on technical reasoning benchmarks in theoretical physics \cite{Chung:2025nsd}. Given the persistence of empirical scaling laws relating performance improvements to increases in model size, data, and training compute \cite{kaplan2020scalinglawsneurallanguage, hoffmann2022trainingcomputeoptimallargelanguage}, the proverbial writing is on the wall. Moreover, there is no significant obstruction to leveraging developments in general-purpose models for problems in high energy physics. Evidence suggests that scaling general-purpose models and improving usage (e.g.~introducing chain of thought reasoning) can surpass the performance of narrowly fine-tuned systems built on smaller base models \cite{wei2023chainofthoughtpromptingelicitsreasoning, singh2025openaigpt5card}. This is known as the ``bitter lesson" of AI: over the long run, methods based on general learning outperform methods that rely heavily on human-engineered structure \cite{sutton2019bitter}. This trend motivates the development of scientific software that will inherit the improvements in underlying general models, rather than designing highly specialized systems that could quickly become obsolete as capabilities advance. On a similar but opposite footing, it is useful to explore whether LLMs can already create useful improvements in the scientific software we currently use even if scaling laws fail sooner than expected.

That being said, benchmark success does not automatically translate into a reliable scientific discovery agent. Between their intrinsically probabilistic nature, propensity for hallucinations, and lack of far-horizon planning, LLMs currently face major challenges in this regard. The effective deployment of current and near-future LLMs in high energy physics stands to benefit from scaffolding reasoning models in a programmatic framework that manages intermediate states, integrates deterministic tools, and enforces constraints or verification.

In this paper we present the first (to our knowledge\footnote{The \fermiacc{} was first presented publicly in \cite{Craig:2026:ReasoningAIccelerators}.}) scaffolded reasoning model designed to
autonomously generate and quantitatively validate plausible hypotheses for high energy physics data: the \fermiacc\footnote{In tribute to the original \texttt{FERMIAC}, an analog Monte Carlo computer. If the modern acronym must stand for something, it might as well be \texttt{F}ast \texttt{E}ngine for \texttt{R}einterpretation: a \texttt{M}achine \texttt{I}ntelligence \texttt{ACC}elerator, but more than anything it honors the idea of a reasoning model for high energy physics as a sort of `AI Fermi'.}, a collection of particle theory agents built on the OpenAI Agents SDK \cite{openai_agents_2026}. The \fermiacc{} is initially tailored to the interpretation of collider data but ultimately generalizable to data from quarks to the cosmos. The core philosophy of the \fermiacc{} is to recast {\it theorists}, rather than data. It reads experimental papers and autonomously produces and validates quantum field theory hypotheses to explain features in these papers using the computational tools available to modern phenomenologists. The \fermiacc{} accomplishes this by coupling commercially-available agents to each other and a deterministic particle physics simulation pipeline. It deploys these agents in a number of novel ways that go beyond orchestrating the simulation pipeline, most notably in (1) an adversarial proposer-critic loop for hypothesis generation and (2) an iterative loop that can feed simulation outcomes back into hypothesis generation. On one hand, our extensive deployment of commercial agents for hypothesis generation embraces the ``bitter lesson'' that general tools scaling with available computational power tend to outperform domain-specific methods, and we expect the \fermiacc{}'s capabilities to improve in step with future advances in reasoning models. On the other hand, our scaffolding with deterministic simulation tools provides robust guardrails against the shortcomings of modern LLMs. 

The adversarial configuration of agents is essential to the effectiveness of the \fermiacc{} at generating novel particle physics hypotheses. While the specific configuration of the \fermiacc{} presented here is designed to reinterpret publicly-available experimental data using fast simulation tools, the core components of the \fermiacc{} may ultimately be put to best use to explore the space of theory hypotheses within experimental collaborations where full simulation tools and correlations are available. More broadly, the viability of \fermiacc{}'s adversarial agent loops with suitably chosen contexts provides a template for pioneering applications of modern reasoning models in high energy theory. 

Needless to say, this work builds on a host of applications of AI in particle physics, which in recent years have evolved from task-specific deep learning to end-to-end, tool-using agentic systems that can co-design simulations and automate analyses. Agents have recently been employed in automating Monte Carlo workflows \cite{AutoFLUKA2024,AutoFLUKA2025}, orchestrating auditable HEP pipelines \cite{HEPTAPOD2025,LLM4HEP2025, Qiu:2026iby}, enabling autonomous \emph{MadGraph} campaigns \cite{MadAgents2026}, supporting experiment-design loops \cite{GRACE2026}, and broadening scientific workflow agency across subfields \cite{ArgoLOOM2025,AgentsOfDiscovery2025, villaescusanavarro2025denarioprojectdeepknowledge}. The \fermiacc{} is a natural extension of these developments into hypothesis generation, and may be coupled to its agentic predecessors to operate simulation tools.

We begin in Sec.~\ref{sec:architecture} by presenting the architecture of the \fermiacc. To demonstrate its effectiveness, in Sec.~\ref{sec:examples} we provide a number of illustrative examples in which the \fermiacc{} delivers and validates novel explanations for mild statistical fluctuations in historical and contemporary LHC publications, beginning with the infamous `750 GeV excess' and continuing on to more recent analyses. We lay out a number of future directions in Sec.~\ref{sec:conc}, and reserve further technical details for a series of appendices. 

\section{\label{sec:architecture} Architecture}

\begin{figure}
    \centering
    \includegraphics[width=0.7\linewidth]{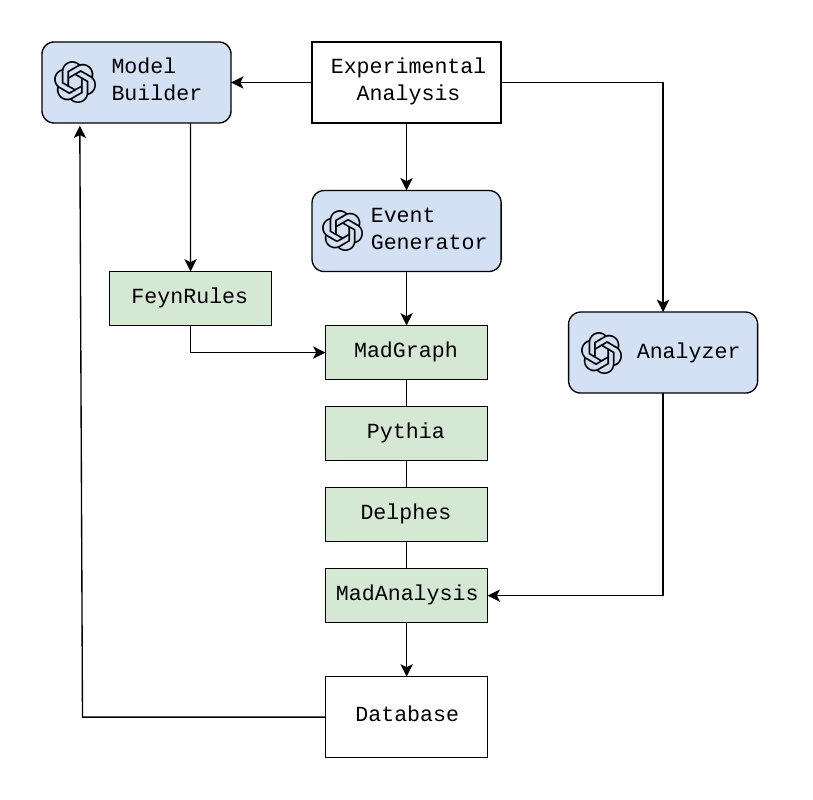}
    \caption{Schematic architecture of the \fermiacc{}. Blue boxes correspond to agent modules. Green shaded boxes correspond to integrated software components.}
    \label{fig:architecture}
\end{figure}

The \fermiacc{} is configured to read an experimental paper\footnote{Here we demonstrate the \fermiacc{} using papers from the ATLAS and CMS collaborations at the Large Hadron Collider, although the deterministic pipeline is equally relevant to any interpretable collider data and the overall framework can be coupled to different pipelines for different datasets in high energy physics.} and deploy a set of scaffolded agents to propose and test hypotheses to explain features in the reported data using deterministic simulation tools. It does so through the following steps, schematically illustrated in Fig.~\ref{fig:architecture}:
\begin{enumerate}
\item Propose and refine field-theoretic hypotheses to explain anomalous features in the data.
\item Encode these hypotheses as extensions of the Standard Model using \emph{FeynRules} \cite{Alloul:2013bka,Christensen:2008bj}, agentically creating the model files while ensuring that they are valid (e.g.~gauge- and Lorentz-invariant) quantum field theories using \emph{FeynRules}' built-in checks.
\item Generate signal events in \emph{MadGraph} \cite{Alwall:2014hca} using topologies from the model hypothesis and simulation parameters extracted from the experimental paper.
\item Pass signal events through decay, showering, and hadronization in \emph{Pythia} \cite{Bierlich:2022pythia83} and perform fast detector simulation in \emph{Delphes} \cite{deFavereau:2013fsa}.
\item Impose selection cuts to obtain signal counts in \emph{MadAnalysis} \cite{Conte:2012fm} and perform rudimentary statistical analysis of the signal significance.
\item Summarize salient details of hypothesis generation, simulation, and signal significance in a database which can be used to seed subsequent hypotheses.
\end{enumerate}

\begin{figure}
    \centering
    \includegraphics[width=0.9\linewidth]{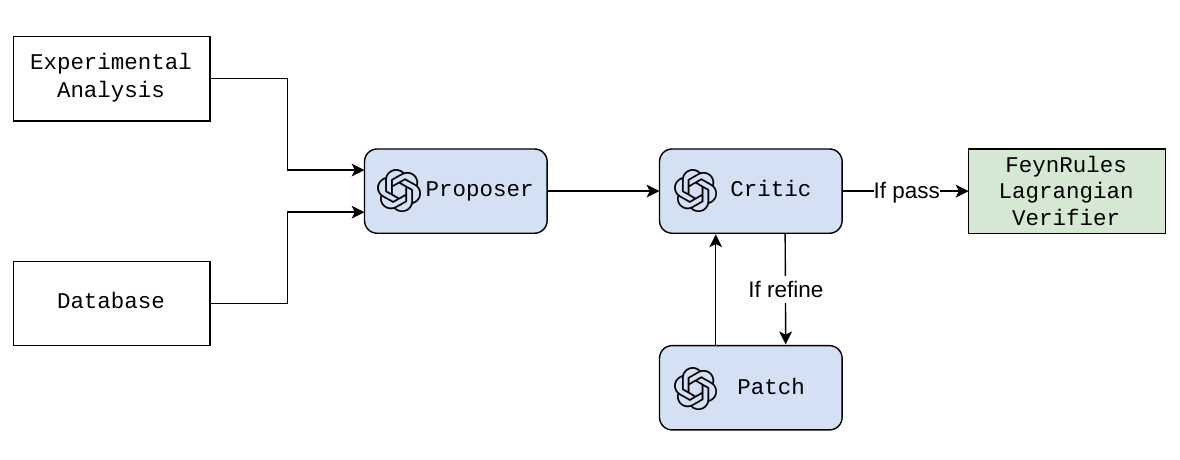}
    \caption{A zoom in on the schematic architecture of the \fermiacc{} model-building module. A proposal agent takes inputs from the experimental analysis and database of previous \fermiacc{} runs and designs a new BSM explanation for a given anomaly. It is then passed between a critic and patching agent until the proposal passes, at which point it is transferred to \emph{FeynRules} for the first layer of verification.}
    \label{fig:modelbuilder}
\end{figure}

To accomplish these tasks, the agents are arranged into three modules:

\begin{itemize}
\item {\bf Model Builder:} Proposes extensions of the Standard Model to explain features in the paper and formalizes them in \emph{FeynRules}. The Model Builder is a multi-agent adversarial refinement loop where a proposer generates hypotheses, a critic provides structured feedback, and a patching agent iterates improvements. Only when the refinement loop converges is the proposal implemented in \emph{FeynRules}. The structure of the Model Builder is schematically illustrated in Fig.~\ref{fig:modelbuilder}. For further details, see Sec.~\ref{sec:modelbuilder}.
\item {\bf Event Generator:} Prepares the fast simulation pipeline (\emph{MadGraph}, \emph{Pythia}, and \emph{Delphes}) to reflect the collider and detector parameters indicated in the paper and orchestrates operation of the pipeline. For further details, see Sec.~\ref{sec:eventgenerator}.
\item {\bf Analyzer:} Identifies and implements the selections defining the paper's signal region(s) in \emph{MadAnalysis} and computes summary statistics for the simulated models. For further details, see Sec.~\ref{sec:analysis-generator}.
\end{itemize}

\subsection{Model Builder \label{sec:modelbuilder}}

The first stage of the \fermiacc{} converts an analysis PDF into a structured beyond-the-Standard-Model (BSM) hypothesis, rather than a free-form textual interpretation. Concretely, each proposal is required to be returned in a typed JSON schema defined using \emph{Pydantic} \cite{pydantic_2024}, a Python library for specifying data structures with explicit types and validation rules. This enforces that the agent returns its output in a well-defined, schema-constrained format that supports deterministic parsing by downstream model-building and simulation tools. In addition to the executable content, the schema also includes structured reasoning notes, enabling subsequent verification and audit of the generated hypothesis. The schema records the analysis signature, paper-grounding anchors, new field content, kinetic and interaction terms, benchmark couplings, order-of-magnitude parameter estimates, collider topology, and a \emph{MadGraph} process configuration.

For each target analysis, the system launches a small ensemble of independent proposal runs with different sampling temperatures. Here the temperature $T$ controls the stochasticity of the language model’s sampling distribution, analogous to thermal fluctuations in statistical mechanics: low $T$ concentrates probability mass on the highest-likelihood completions, while higher $T$ explores the Boltzmann tail of the distribution and increases the probability of exploring more novel hypotheses. This provides a controlled way to probe the entirety of theory space.

A central design choice is the separation between the \emph{full model story} and the \emph{collider benchmark} that is actually passed to downstream model-building tools. This is necessary because the present UFO/\emph{FeynRules} path is currently restricted to pre-electroweak-symmetry-breaking, gauge-basis extensions of the Standard Model, and does not support new gauge sectors, additional vacuum expectation values, explicit broken-phase operators, or custom mass-diagonalization metadata. Each generated hypothesis therefore includes an explicit contract for the executable benchmark, specifying which fields and couplings must survive into the UFO, which couplings are optional, and which ultraviolet features are acknowledged but intentionally left outside the executable realization.\footnote{At present, one shortcoming of our software is our ability to automatically verify the mapping from the UV theory to the executable benchmark. Future integration with software such as SARAH \cite{Staub:2013tta} will allow us to match features between UV models and UFO implementations.}

In order to ensure novelty of model-building proposals, we mandate retrieval against a proposal database. Rather than comparing raw model names, the system first performs a coarse search over field content: each new field is labeled by spin class, CP label, color representation class, electroweak representation class, hypercharge, and self-conjugacy, and this representation is used to retrieve structurally similar prior proposals. The retrieved records include not only field content, but also interaction terms, kinetic terms, topology summaries, \emph{MadGraph} process strings, and previous simulation outcomes. Final identification of equivalence is then determined at a finer level, requiring consistency in the physical structure—including production mode, decay chain and interaction pattern—rather than relying on field content alone. Novelty is thus assessed at the level of physical structure.

Each candidate model then enters an iterative propose--critique--patch loop. The critic evaluates six dimensions: analysis grounding, novelty with respect to the paper and proposal database, physical consistency, specificity, compatibility with the UFO pipeline, and parameter estimation. In particular, every new field mass and every coupling must be accompanied by an explicit order-of-magnitude justification, so that benchmark values are derived from phenomenological reasoning rather than chosen ad hoc. The critic assigns \texttt{PASS}, \texttt{REFINE}, or \texttt{FAIL} to each category, and the overall decision is set by the most severe category. When the proposal is repairable, a patching agent applies only the minimal requested modifications and the loop repeats until the hypothesis passes, fails irreparably, or reaches the configured patch limit. All intermediate proposals, critiques, and patches are logged, while the final proposal summary is stored together with its temperature and critique decision and inserted into the searchable proposal database for reuse in later runs. This yields an ensemble of BSM hypotheses that are already constrained by downstream executability requirements.

Further technical details are provided in App.~\ref{app:hypothesis_impl}.

\subsection{From Hypothesis to Simulated Events \label{sec:eventgenerator}}

Once a hypothesis has been approved, the next part of the pipeline turns it into an executable collider benchmark and then into simulated events. This stage is deliberately conservative. Rather than trying to support every possible BSM construction, it uses a restricted workflow with a number of guardrails so that only models that can be written, exported, and run reliably are passed downstream.

The first step is to prepare the event-generation settings. A small helper reads the analysis PDF together with a short summary of the selected hypothesis and proposes a minimal \emph{MadGraph} run card, including the beam energy, event count, and a few simple object-level cuts when these are clearly stated in the paper. This keeps the generation settings tied to the experimental analysis, while the particle content and interactions remain part of the model definition itself.

The model-building step then converts the approved hypothesis into a \emph{FeynRules} extension and exports it as a UFO model. Here the pipeline adds concrete PDG codes and parameter-block assignments for the new particles and couplings, and constructs the smallest benchmark model needed to realize the requested process. This step is surrounded by guardrails. In particular, the benchmark must remain in the unbroken Standard Model gauge basis: the pipeline does not support new gauge bosons, extra vacuum expectation values, explicit broken-phase fields such as $A$, $Z$, and $W$, or custom mass-mixing and diagonalization machinery. It also rejects several patterns that are known to cause failures in this workflow, such as malformed Higgs bilinears or unsafe SU(2) field-strength operators. If the generated extension passes these checks, the pipeline runs \texttt{LoadModel}, \texttt{CheckHermiticity}, and \texttt{WriteUFO}, sanitizes the exported UFO for \emph{MadGraph}, checks that the benchmark couplings requested by the hypothesis actually appear in the exported model, and finally performs a minimal \emph{MadGraph} \texttt{generate} test to verify that the requested hard process has at least one diagram. If a stage fails before yielding a usable validated UFO, the corresponding diagnostics are fed back into a repair loop and the \emph{FeynRules} extension is rewritten.

After a UFO model has passed these guardrails, the pipeline builds the event-generation job itself. A template \emph{MadGraph} configuration is combined with the hypothesis-level process definition, the run-card settings inferred from the paper, and the benchmark parameter values extracted from the UFO build. These parameter values are written into the \emph{MadGraph} \texttt{param\_card} explicitly, so that the event-generation step uses the same benchmark that passed the UFO validation stage. \emph{MadGraph} then generates the requested hard process and, in the default configuration, passes the events to \emph{Pythia8} for parton showering. Detector simulation is done independently via \emph{Delphes} with the standard experiment dependent card, and lastly the main reinterpretation downstream using  \emph{MadAnalysis}. The same machinery can be used either for a single benchmark-point sample or for a capped parameter scan over the suggested benchmark ranges.

Further technical details are provided in App.~\ref{app:ufo_mg5_pipeline}.

\subsection{From Simulated Events to Hypothesis Testing \label{sec:analysis-generator}}

Finally, we implement an agentic pipeline to convert the signal selection in the analysis PDF into an executable reinterpretation package. The system takes the analysis PDF as input and produces (i) a structured machine-readable representation of the analysis, (ii) a \emph{MadAnalysis~5} implementation, and (iii) statistical outputs suitable for recasting. The design goal is to reduce the amount of hand-written analysis code while preserving a transparent intermediate representation that can be inspected, corrected, and archived.

\begin{figure}
    \centering
    \includegraphics[width=0.9\linewidth]{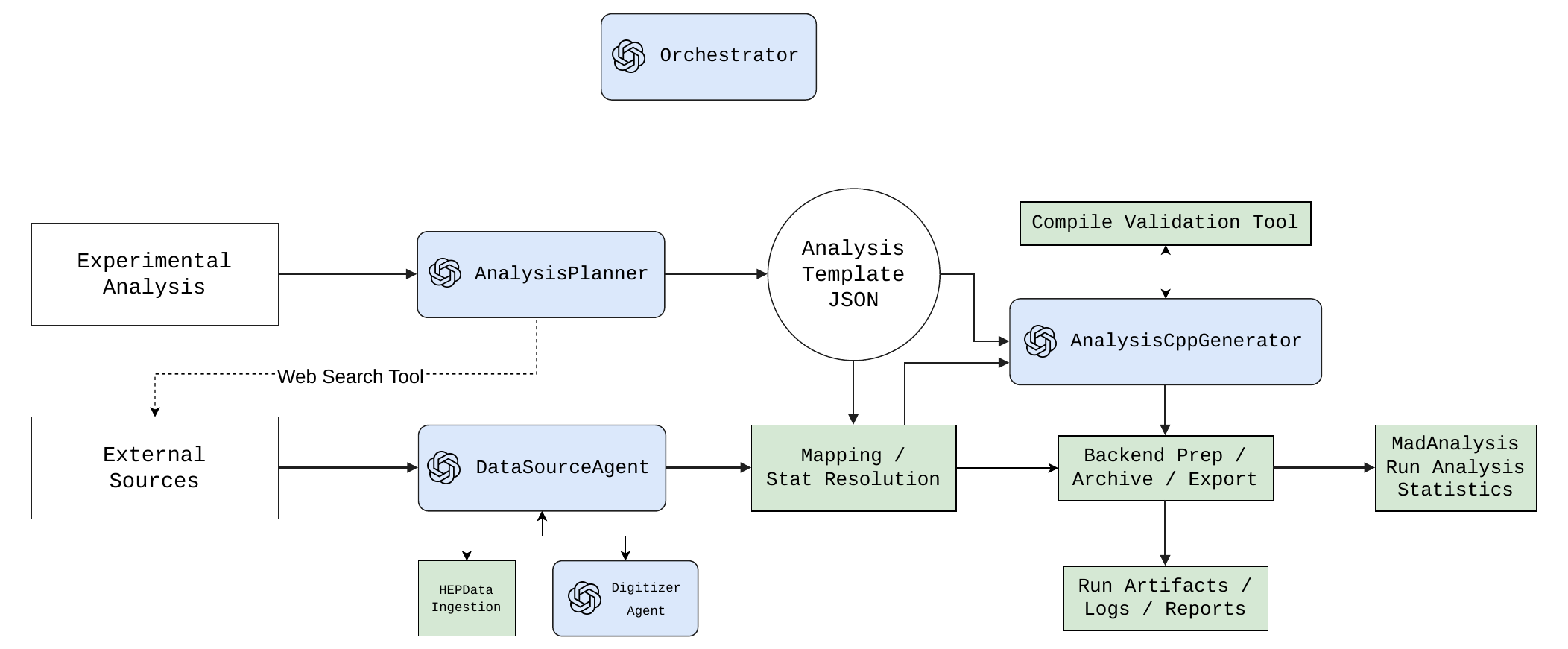      }
    \caption{A zoom in on the schematic architecture of the \fermiacc{} analysis generator module.}
    \label{fig:analysis_generator}
\end{figure}

At a high level, the pipeline performs four tasks:
\begin{enumerate}
    \item extract the analysis logic from the source document,
    \item resolve the numerical ingredients needed for validation and statistics,
    \item generate an executable analysis implementation,
    \item run the analysis and compute exclusions.
\end{enumerate}

A central design choice is to represent each analysis explicitly in structured form before code generation. Rather than directly prompting a model for final C++ code, the system first builds a normalized analysis template containing the signal regions, control or validation regions, cutflow structure, observables, and statistical inputs. This intermediate representation acts as the contract between the semantic extraction stage and the code-generation stage. It also provides a natural place to archive provenance and to validate consistency before execution.

The pipeline is organized as a sequence of specialized agents with distinct responsibilities:
\begin{itemize}
    \item \textbf{Analysis planner:} reads the source material and reconstructs the semantic structure of the search, including object definitions, selection logic, region hierarchy, and region families.
    \item \textbf{Data-source agent:} resolves numerical inputs such as observed counts, background predictions, uncertainties, and auxiliary tables using structured sources (e.g.\ HEPData), extracted tables, or digitized information from plots when needed.
    \item \textbf{Code-generation agent:} converts the structured analysis into a \emph{MadAnalysis~5} implementation, including the cutflow logic, region definitions, and analysis metadata required by the recasting framework.
\end{itemize}
This separation is important in practice. It allows semantic interpretation, numerical data acquisition, and executable code synthesis to be improved independently, and it makes failure modes easier to diagnose than in a single end-to-end prompt.

Once generated, the analysis is exported into a \emph{MadAnalysis~5} installation and compiled as a standard user analysis in expert mode. The system then runs the analysis on detector-level event samples and stages the resulting outputs for statistical interpretation. When supported, exclusion limits are computed using the native \emph{MadAnalysis~5} recasting machinery rather than an external post-processing layer. This keeps the statistical evaluation close to the runtime that produced the cutflows and reduces opportunities for mismatches between the event-selection output and the statistical input.

Each run is archived with a timestamped run identifier together with the generated code, metadata, tool outputs, and execution artifacts. In practice this means that a given analysis can be traced from the original source document through planner output, numerical-input resolution, generated \emph{MadAnalysis~5} files, cutflows, and final statistical summaries. This archival structure is especially useful for iterative debugging, regression testing, and comparison between multiple generated versions of the same analysis.

The pipeline performs consistency checks at multiple stages. In particular, it verifies agreement between the generated analysis code, the declared region metadata, and the statistical inputs used for recasting. This is important because many analyses mix different region topologies, such as disjoint histogram bins, cumulative threshold scans, and control-region families. By validating the generated topology before execution, the system can catch common failure modes such as mismatched region identifiers or incomplete statistical metadata.

The system is intended as an analysis-construction framework rather than a single monolithic model call. Its main contribution is the explicit combination of structured reasoning, tool use, and executable code generation. In this sense, the workflow is agentic in a practical rather than purely rhetorical sense: different stages maintain their own state, produce inspectable outputs, and can trigger targeted recovery steps when inconsistencies are found. This modularity has proved especially useful for analyses whose natural region structure is not a flat list of independent signal regions, but rather a family of related thresholds or category-dependent selections.

A key desirable feature that is left to future work is a physics validation layer, which will reproduce signal efficiencies for a model provided in the analysis and compare them with the published efficiency. This will serve both as a test but also provide options for reweighting efficiencies to produce higher fidelity recasts.

Further technical details are provided in App.~\ref{app:agentic-recasting-pipeline}.

\section{\label{sec:examples} Examples}

Here we present several examples of LHC analyses with mild statistical fluctuations to demonstrate the efficacy of the \fermiacc{}. For the sake of illustration, we begin with the most famous statistical fluctuation of the past decade, the `750 GeV excess' observed in early Run 2 data. While the extraordinary amount of attention devoted to the 750 GeV excess means the space of signal explanations is already spanned in the agents' training data, it nonetheless provides a valuable benchmark against which to judge the \fermiacc{}'s capabilities. We then consider several more recent, interpretable analyses with mild fluctuations. As these have received a fraction of the attention devoted to the 750 GeV excess, the full space of signal explanations is unlikely to exist in the agents' training data and therefore provides a more realistic indication of the \fermiacc{}'s potential to explain future anomalies. It bears emphasizing that these proposals were not run through an iterative loop for refinement, which illustrates the \fermiacc{}'s first-pass potential at hypothesis generation. 

\subsection{CMS Diphoton Resonance: the `750 GeV Excess'}

The CMS ``Search for new physics in high mass diphoton events in proton-proton collisions at 13 TeV'' \cite{CMS:2015dxe} Physics Analysis Summary reported a 2.6$\sigma$ (local) excess in the diphoton final state around $m_{\gamma \gamma} = 760$ GeV, which alongside its ATLAS companion (see below) gave rise to more than 500 model-building papers. In the PAS note, the excess is interpreted in terms of a Randall-Sundrum Kaluza-Klein graviton (spin-2) resonantly produced in gluon fusion and decaying to two photons.

The \fermiacc{} produced and tested a number of viable proposals at the level of the PAS note, which together span the majority of models proposed at the time of the 750 GeV anomaly; a representative example is reproduced below. 

\subsubsection{Proposal \texttt{544c3421}: Pseudoscalar coupled to vector-like quarks}

\begin{figure}
    \centering
    \includegraphics[width=0.7\linewidth]{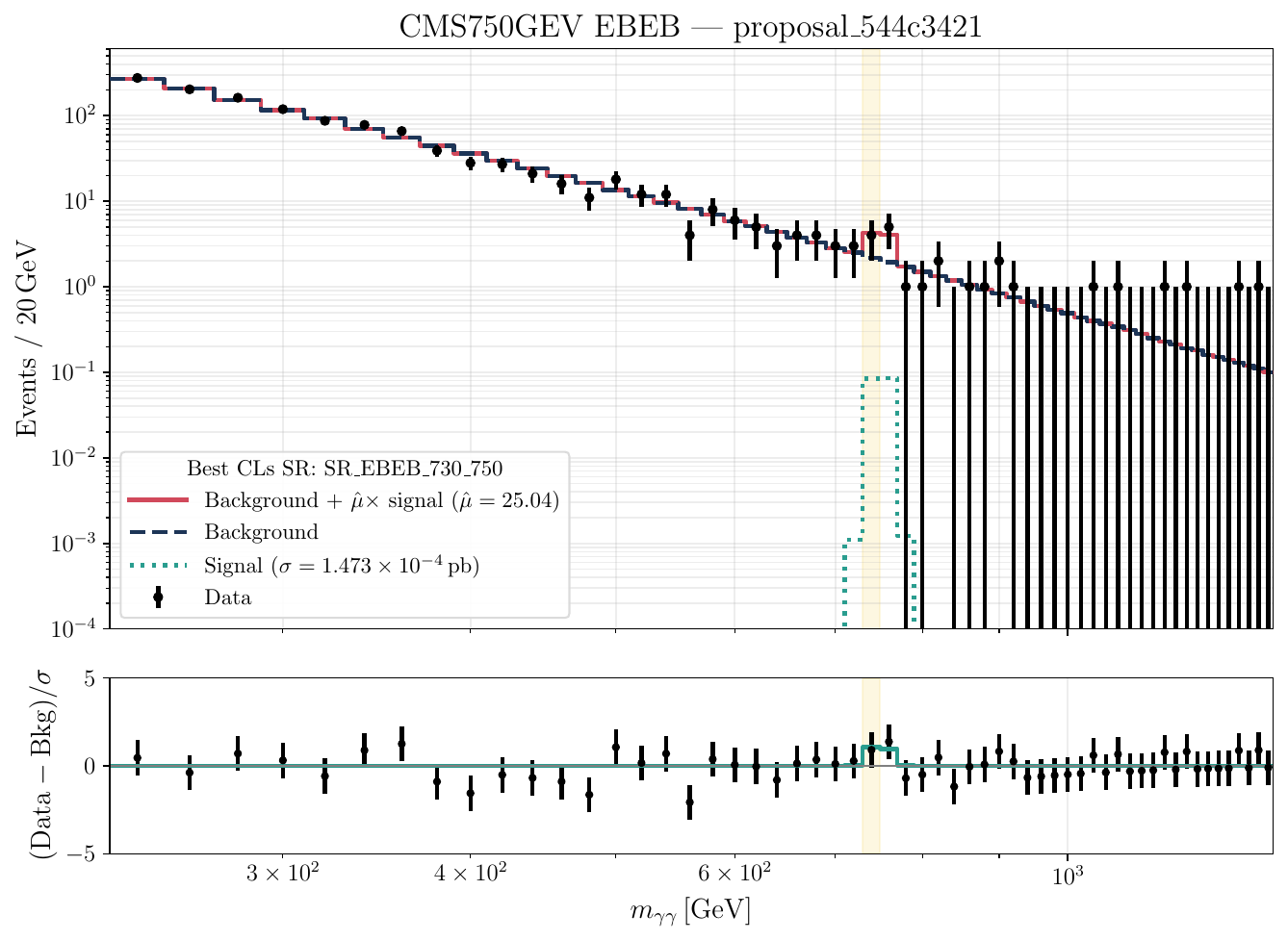}
    \includegraphics[width=0.7\linewidth]{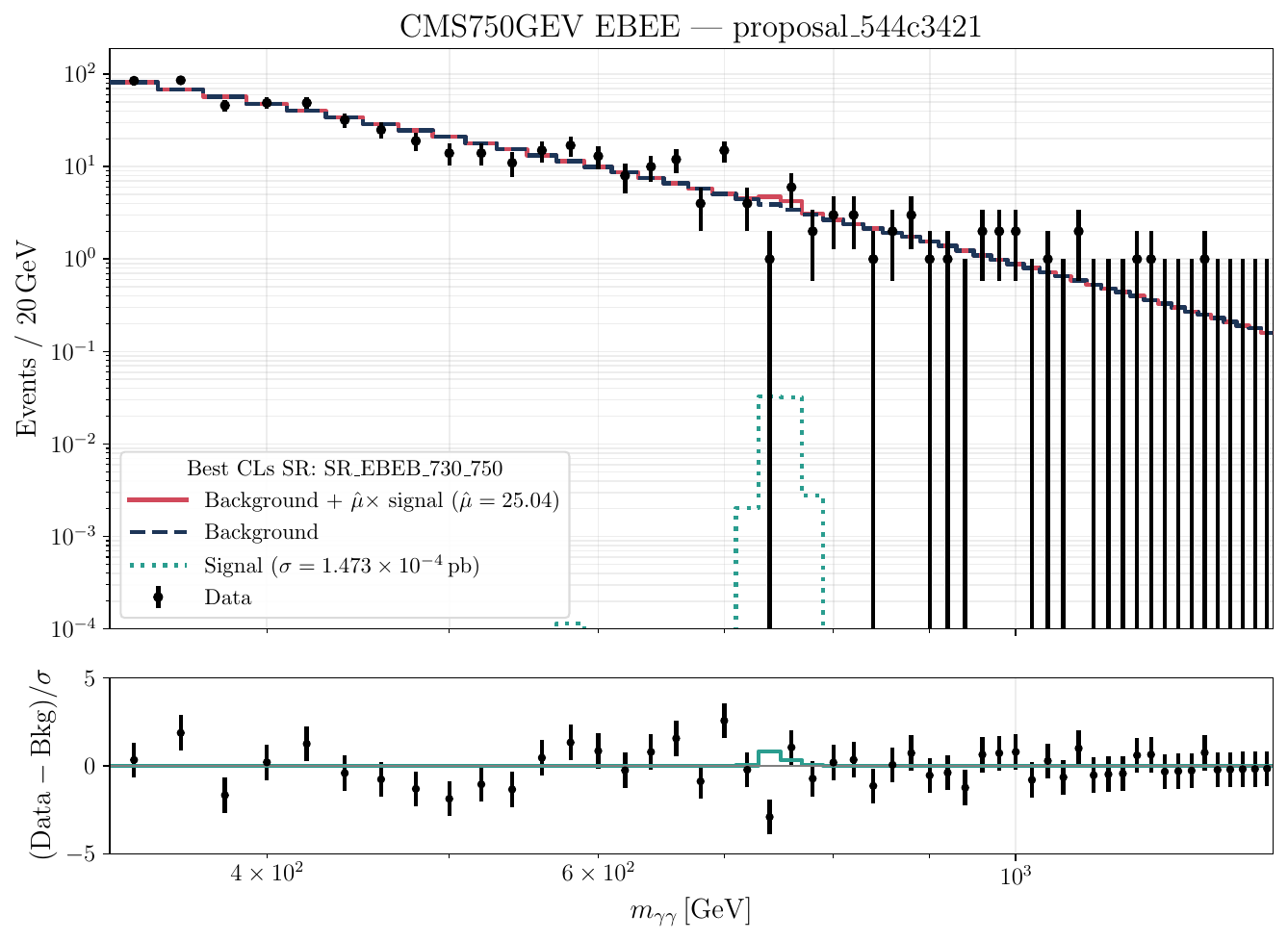}
    \caption{Background prediction and observed counts from the CMS diphoton search \cite{CMS:2015dxe} in the EBEB (top) and EBEE (bottom) channels overlaid with the simulated best-fit signal counts from the \fermiacc{}'s proposal \texttt{544c3421} for a pseudoscalar coupled to heavy vector-like quarks.}
    \label{fig:proposal544c3421}
\end{figure}

This proposal involves a Standard Model-neutral 750 GeV pseudoscalar $A$ with a Yukawa coupling to 1.2 TeV vector-like quarks $T,\bar{T}$ charged under $SU(3)_c$ and $U(1)_Y$ but not $SU(2)_L$ in order to generate EFT couplings $A \, G^{a, \mu \nu} \tilde G^a_{\mu \nu}$ and $A \, B^{\mu \nu} \tilde B_{\mu \nu}$ at low energies. (Many variations of this model appeared during the initial furore surrounding the 750 GeV excess.) The pseudoscalar mass is chosen to reproduce the observed bump, while the quark mass is chosen to close the $A \rightarrow T \bar T$ channel while giving plausible EFT couplings. The Yukawa coupling was deliberately chosen near but below the border of perturbative unitarity. The \fermiacc{} stated the one-loop Wilson coefficients of the EFT operators $A G^{a, \mu \nu} \tilde G^a_{\mu \nu}$ and $A B^{\mu \nu} \tilde B_{\mu \nu}$ obtained by integrating out the vector-like quarks, checked that its parameter choices led to a reasonable production rate, significant branching ratio into diphotons, and narrow width, and implemented the EFT operators directly in the UFO file. It then pushed this proposal through the deterministic simulation pipeline to produce the histograms for both analysis regions shown in Fig.~\ref{fig:proposal544c3421}.

\subsection{ATLAS Diphoton Resonance: the `750 GeV Excess'}
The ATLAS ``Search for resonances decaying to photon pairs in 3.2 fb$^{-1}$ of $pp$ collisions at $\sqrt{s}= 13$ TeV with the ATLAS detector'' conference note \cite{TheATLAScollaboration:2015mdt} reported a 3.6$\sigma$ (local) excess at $m_{\gamma \gamma} = 750$ GeV with indications of a significant width $\sim 45$ GeV. The conference note includes a signal topology for production of a resonance in gluon fusion with a decay to two photons, but does not include a signal model interpretation.

The \fermiacc{} produced and tested a number of viable proposals at the level of the conference note, of which two examples are reproduced below. Interestingly, some of its proposals for the ATLAS diphoton excess differed from CMS counterparts due to specific features in the ATLAS analysis.

\subsubsection{Proposal \texttt{5e41fd9e}: Hypercharge axion}

\begin{figure}
    \centering
    \includegraphics[width=0.7\linewidth]{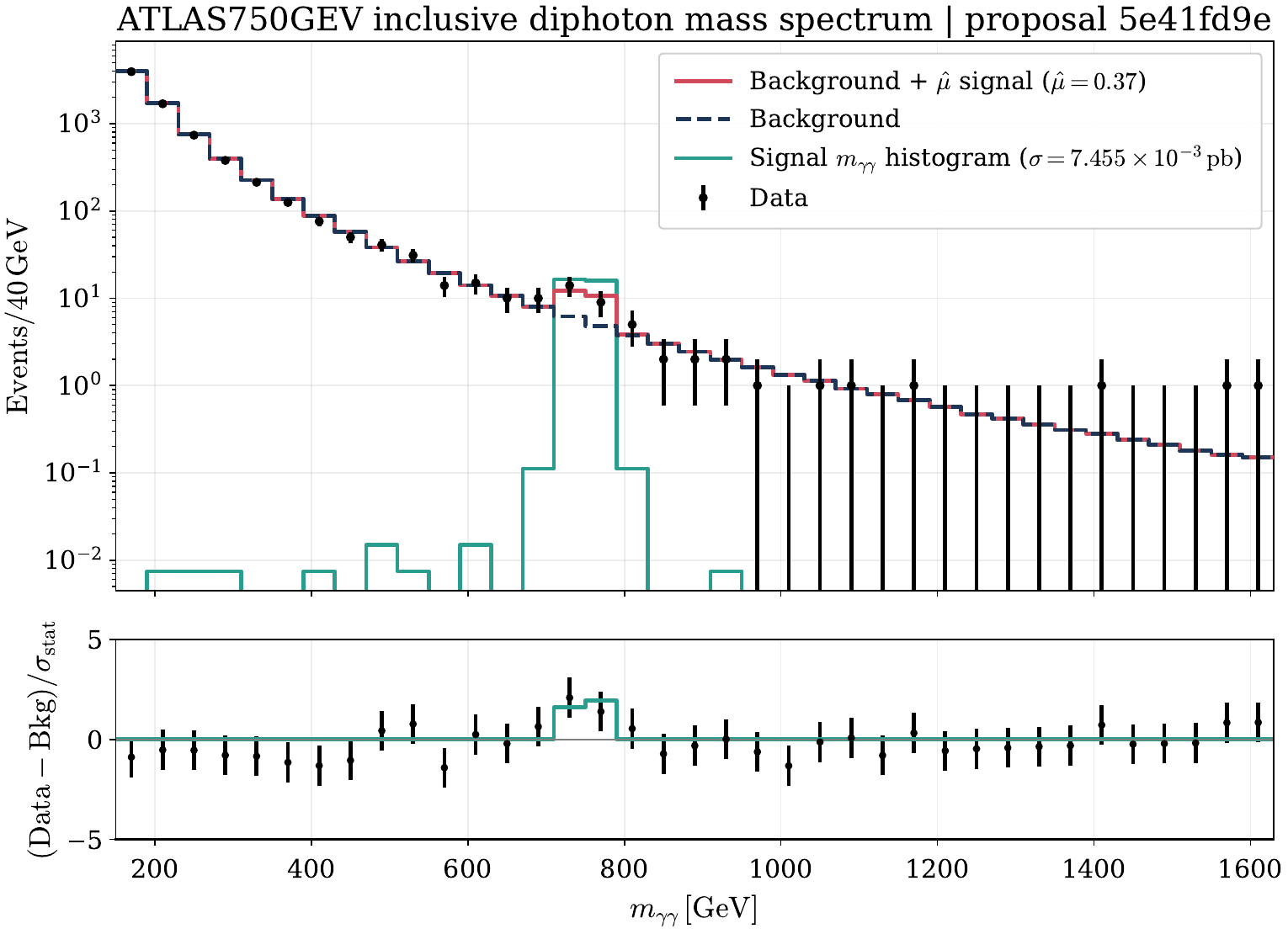}
    \caption{Background prediction and observed counts from the ATLAS diphoton search \cite{TheATLAScollaboration:2015mdt} overlaid with the simulated best-fit signal counts from the \fermiacc{}'s proposal \texttt{5e41fd9e} for a hypercharge axion.}
    \label{fig:proposal5e41fd9e}
\end{figure}

This proposal involves a Standard Model-neutral 750 GeV pseudoscalar $A$ with only the EFT coupling $A \, B^{\mu \nu} \tilde B_{\mu \nu}$ to the Standard Model hypercharge gauge field. (This particular highly-minimal model was not among the most common explanations for the 750 GeV, but did appear in \cite{Ben-Dayan:2016gxw}.) The pseudoscalar mass is chosen to reproduce the observed bump, while the EFT coupling is chosen to give a plausible production rate in vector boson fusion $pp \rightarrow A + jj$ and branching ratio $A \rightarrow \gamma \gamma$. In proposing this hypothesis, the \fermiacc{} noted that the two forward quark jets implied by the VBF topology are neither required nor vetoed by the inclusive ATLAS search. It then pushed this proposal through the deterministic simulation pipeline to produce the histogram shown in Fig.~\ref{fig:proposal5e41fd9e}.

\subsubsection{Proposal \texttt{7a3fb5a1}: Scalar with EFT couplings}

\begin{figure}
    \centering
    \includegraphics[width=0.7\linewidth]{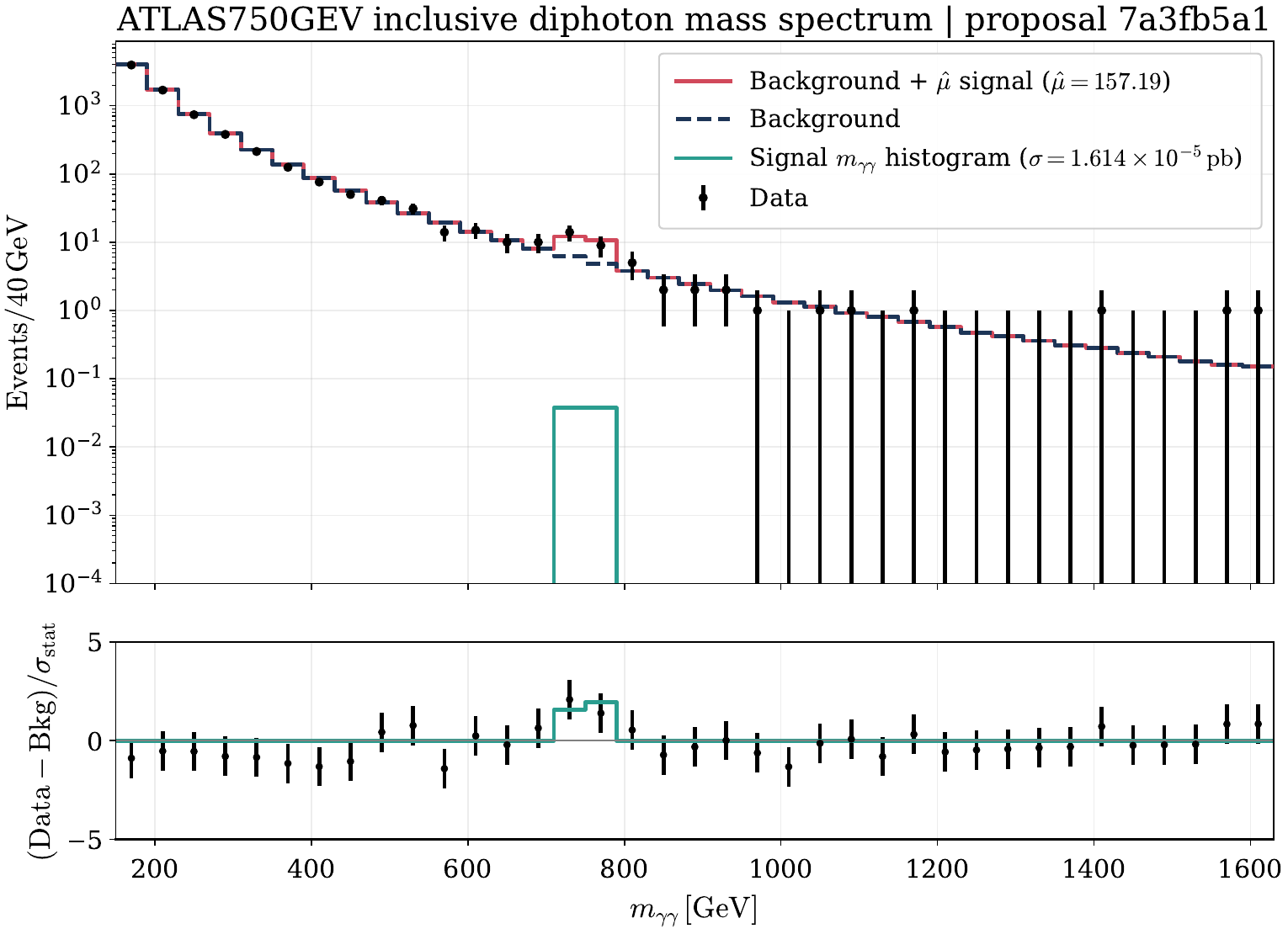}
    \caption{Background prediction and observed counts from the ATLAS diphoton search \cite{TheATLAScollaboration:2015mdt} overlaid with the simulated best-fit signal counts from the \fermiacc{}'s proposal \texttt{7a3fb5a1} for a scalar with EFT couplings. The small signal cross section is a consequence of the \fermiacc{} choosing a small gluon coupling to preserve the diphoton branching ratio.}
    \label{fig:proposal7a3fb5a1}
\end{figure}

This proposal involves a Standard Model-neutral 750 GeV scalar $S$ with EFT couplings $S \, G^{a, \mu \nu} G^a_{\mu \nu}$ and $S \, B^{\mu \nu} B_{\mu \nu}$. (Many variations of this model appeared during the initial furore surrounding the 750 GeV excess.) The scalar mass is chosen to reproduce the observed bump, while the EFT couplings are chosen to give a predominant branching ratio $S \rightarrow \gamma \gamma$. (A notable failure mode of the initial proposal is that it chose a small coupling to gluons in order to avoid suppressing the diphoton branching ratio, which led to an insufficient signal cross section.) The \fermiacc{} then pushed this proposal through the deterministic simulation pipeline to produce the histogram shown in Fig.~\ref{fig:proposal7a3fb5a1}.

\subsection{CMS Paired Dijet Resonances}

The CMS ``Search for resonant production of pairs of dijet resonances through broad mediators in proton-proton collisions at $\sqrt{s} = 13$ TeV'' paper \cite{CMS:2025hpa} reported a 3.9$\sigma$ (local) excess in four-jet final states with features in both the four-jet invariant mass $m_{4j}$ and paired dijet invariant mass $m_{2j}$ corresponding to $(m_{4j},m_{2j})= (8.6,2.15)$ TeV. A second 3.9$\sigma$ (local) excess is observed at $(m_{4j},m_{2j})= (3.6,1.0)$ TeV. In the paper, the excesses are interpreted in terms of a color sextet scalar diquark $S_{uu}$ or $S_{dd}$ (charge $4/3$ or $-2/3$) decaying to pairs of vector-like quarks, which in turn decay to quarks and gluons to give the dijets. As a relatively recent excess, it has only been subject to three proposed explanations \cite{Bittar:2025rcw, Costache:2025bjc, Dobrescu:2025hyv}.

The \fermiacc{} produced novel candidate explanations for both excesses. A representative explanation for each of the two excesses is summarized below. These model explanations are distinct from those in the original CMS paper \cite{CMS:2025hpa} {\it and} the subsequent theory reinterpretations. A notable shortcoming of both models is that the \fermiacc{} proposed an overly conservative coupling of the parent scalar to gluons in order to preserve the decay into pairs of the daughter particles. Operating the \fermiacc{} in an iterative loop could balance the production and decay constraints within the model.

\subsubsection{Proposal \texttt{449ba23a}: Singlet plus octet scalars}

\begin{figure}
    \centering
    \includegraphics[width=0.7\linewidth]{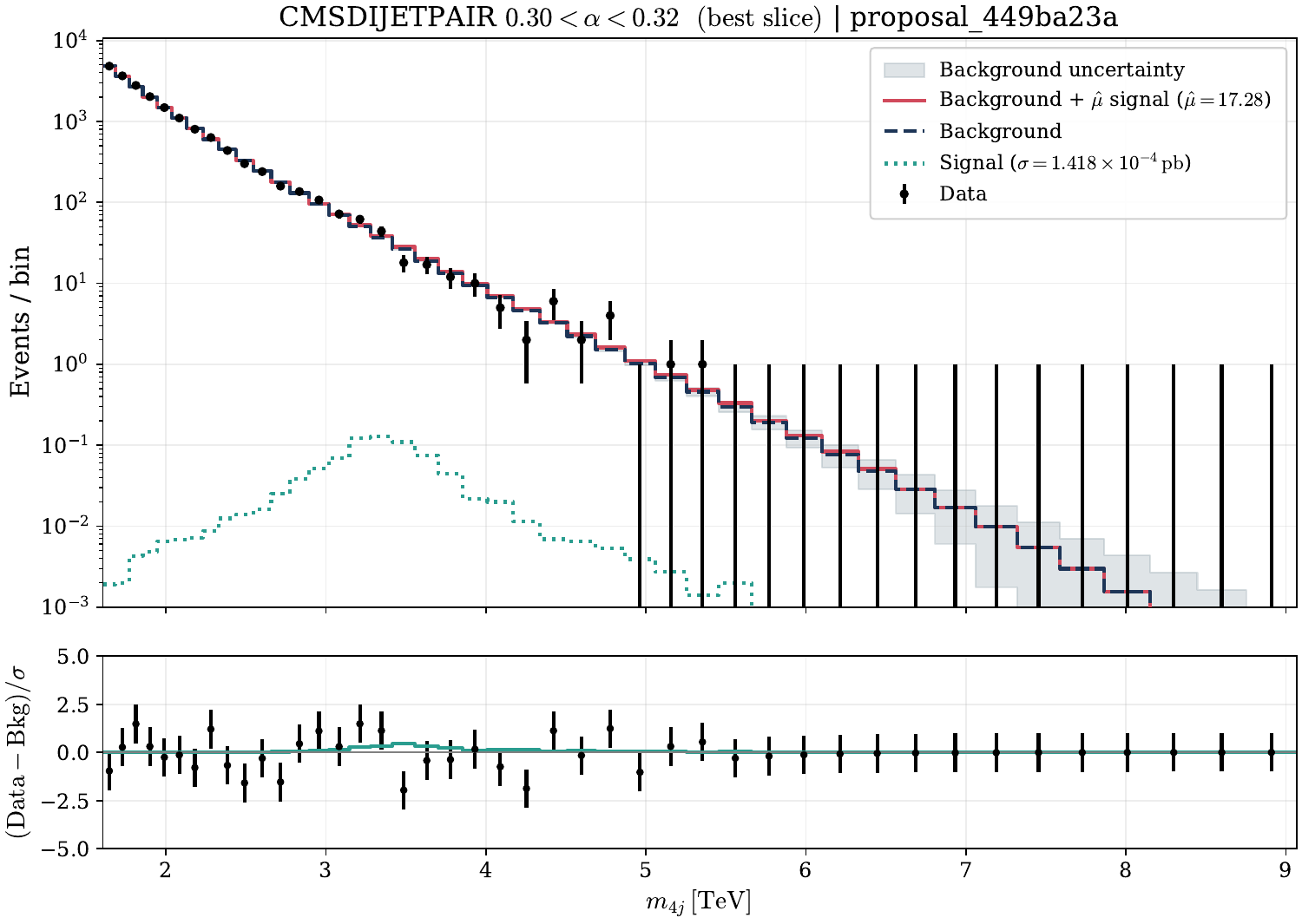}
    \caption{Background prediction and observed counts from the relevant slice of the CMS dijet pair search \cite{CMS:2025hpa} overlaid with the simulated best-fit signal counts from the \fermiacc{}'s proposal \texttt{449ba23a} for singlet plus octet scalars. The small signal cross section is a consequence of the \fermiacc{} choosing a small gluon coupling to preserve the singlet scalar branching ratio into octet scalars.}
    \label{fig:proposal449ba23a}
\end{figure}

This proposal involves a Standard Model-neutral 3.6 TeV scalar $S$ coupled to a 1 TeV $SU(3)_c$ octet scalar $O^a$ via a cubic $S O^a O^a$ interaction. Both couple singly to gluons via the EFT operators $S \, G^{a,\mu \nu} G^a_{\mu \nu}$ and $d_{abc} O^a G^{b,\mu \nu} G^c_{\mu \nu}$, in addition to the gluonic couplings of $O^a$ implied by its color charge. The masses are chosen to reproduce the second $3.9\sigma$ excess, while the trilinear coupling is taken to be large in order to reproduce the broad width of the parent particle. (A notable failure mode of the initial proposal is that it chose a small coupling of $S$ to gluons in order to avoid suppressing the $S \rightarrow OO$ branching ratio, leading to an insufficient signal cross section.) The \fermiacc{} then pushed this proposal through the deterministic simulation pipeline to produce the histogram shown in Fig.~\ref{fig:proposal449ba23a}.

\subsubsection{Proposal \texttt{896abfa5}: Singlet scalar plus octet pseudoscalar}

\begin{figure}
    \centering
    \includegraphics[width=0.7\linewidth]{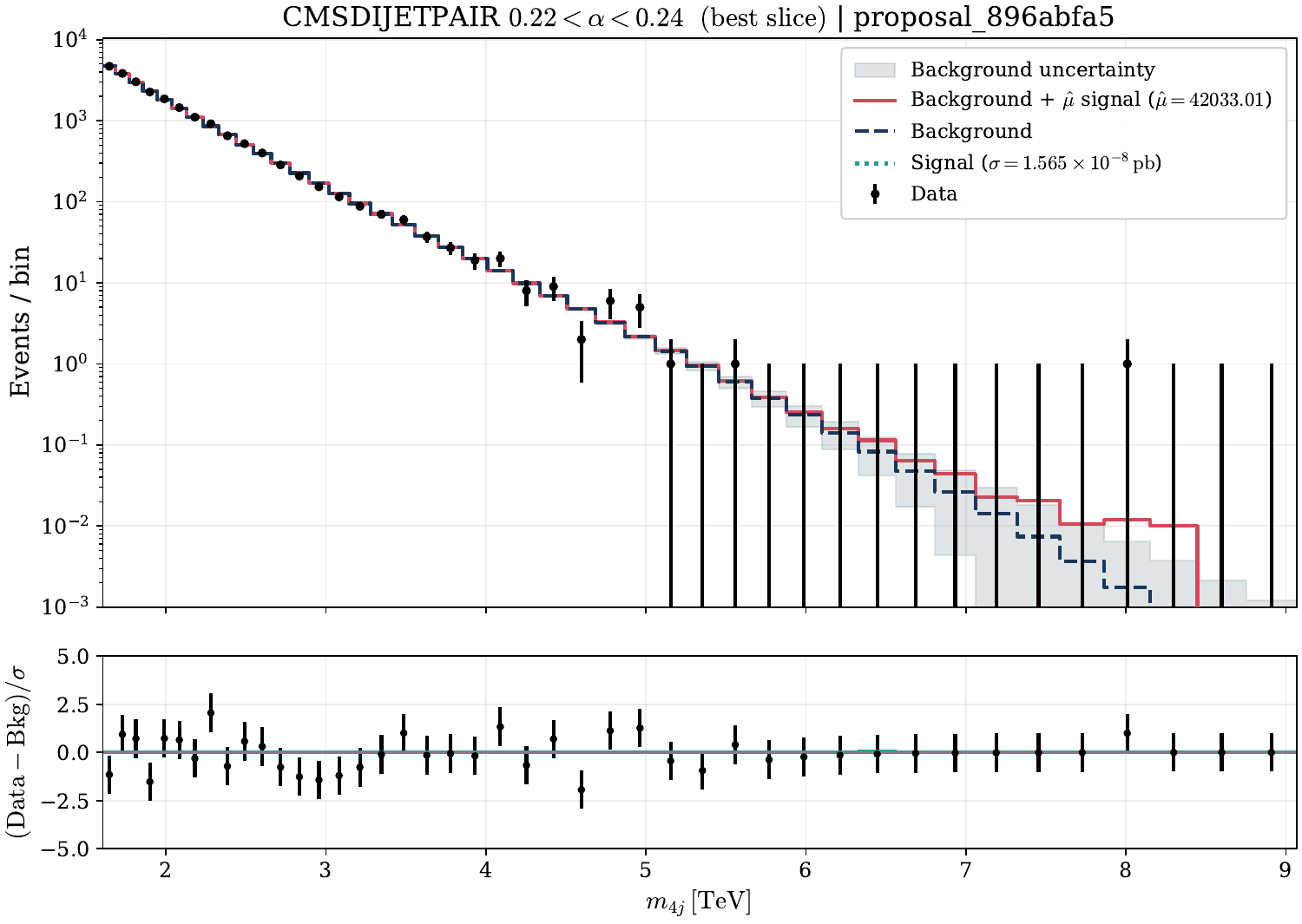}
    \caption{Background prediction and observed counts from the relevant slice of the CMS dijet pair search \cite{CMS:2025hpa} overlaid with the simulated best-fit signal counts from the \fermiacc{}'s proposal \texttt{896abfa5} for a singlet scalar plus octet pseudoscalar. The small signal cross section is a consequence of the \fermiacc{} choosing a very small gluon coupling to preserve the singlet scalar branching ratio into octet pseudoscalars.}
    \label{fig:proposal896abfa5}
\end{figure}

This proposal involves a Standard Model-neutral 8.6 TeV scalar $S$ coupled to a 2.15 TeV $SU(3)_c$ octet pseudoscalar $P^a$ via a cubic $S P^a P^a$ interaction. Both couple singly to gluons via the EFT operators $S \, G^{a,\mu \nu} G^a_{\mu \nu}$ and $d_{abc} P^a G^{b,\mu \nu} \tilde G^c_{\mu \nu}$, in addition to the gluonic couplings of $P^a$ implied by its color charge. The masses are chosen to reproduce the first $3.9\sigma$ excess, while the trilinear coupling is taken to be large in order to reproduce the broad width of the parent particle. (A notable failure mode of the initial proposal is that it chose a small coupling of $S$ to gluons in order to avoid suppressing the $S \rightarrow PP$ branching ratio, leading to an insufficient signal cross section.) The \fermiacc{} then pushed this proposal through the deterministic simulation pipeline to produce the histogram shown in Fig.~\ref{fig:proposal896abfa5}.

\subsection{ATLAS Dijet Resonance plus Lepton}

The ATLAS ``Search for dijet resonances in events with an isolated charged lepton using $\sqrt{s} = 13$ TeV proton–proton
collision data collected by the ATLAS detector'' paper \cite{ATLAS:2020zzb} reported a 2.8$\sigma$ (local) excess in dijet plus single isolated electron final state corresponding to $m_{jj} = 1.3$ TeV. In the paper, the excess is interpreted in four different models: a technicolor model with an electroweak triplet vector $\rho$ decaying to a $W$ boson and a techni-pion, where the $W$ gives the lepton and the techni-pion the dijet; a $W'$ and $Z'$ model where $W' \rightarrow W (Z' \rightarrow jj)$; a singly-charged Higgs boson produced in $tb$ associated production with a $tb$ decay, hence a $t \bar{t} b \bar{b}$ final state; and a $Z'$ model where t-channel quark exchange gives a $W/Z$ for the lepton plus dijets from $Z' \rightarrow jj$.

The \fermiacc{} produced a novel candidate explanation for the observed excess, a non-trivial task given the four distinct model interpretations already contained in the experimental paper.

\subsubsection{Proposal \texttt{f597e791}: Octet scalar with EFT couplings}

\begin{figure}
    \centering
    \includegraphics[width=0.7\linewidth]{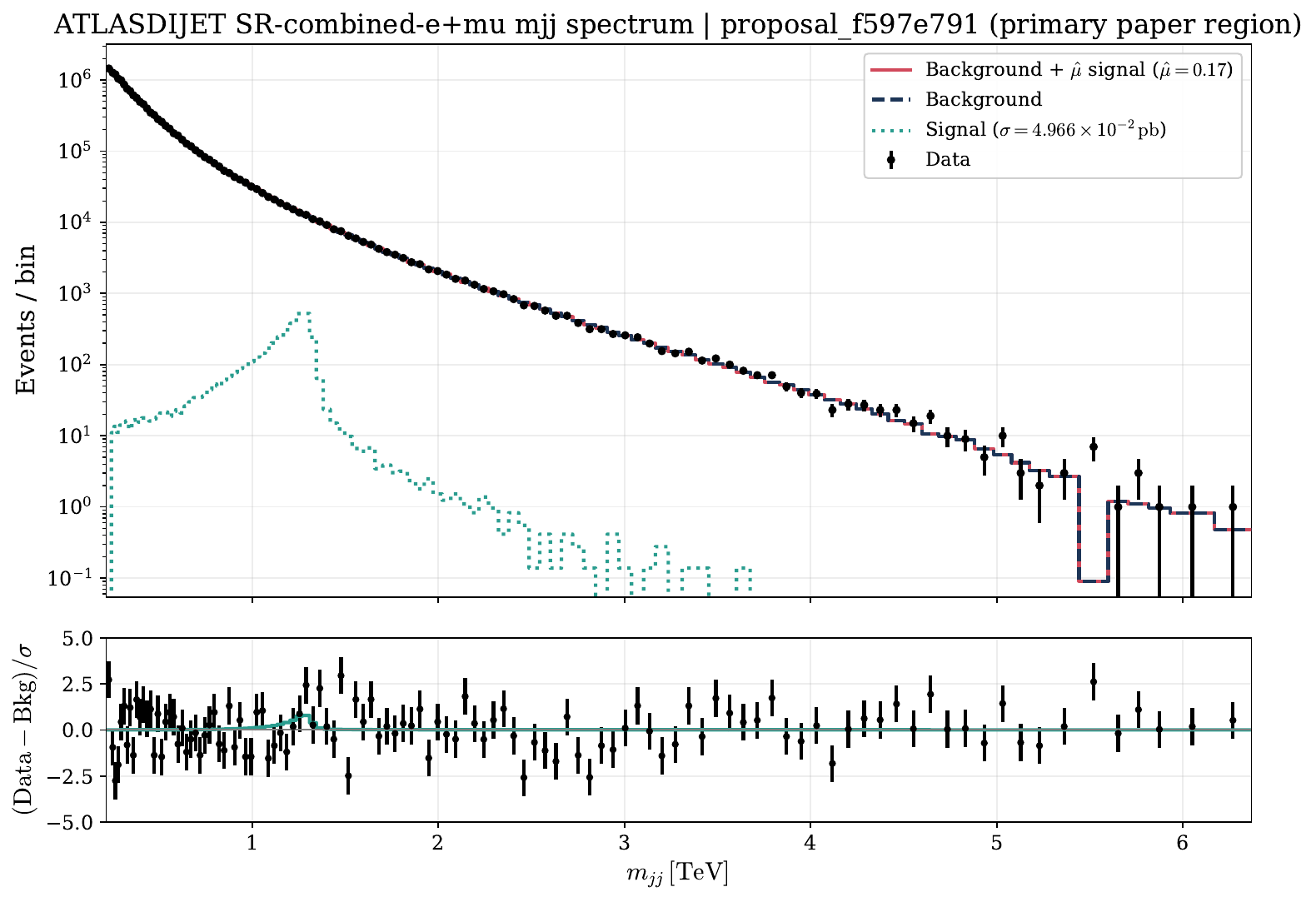}
    \caption{Background prediction and observed counts from a relevant channel of the ATLAS  dijet resonance plus lepton search \cite{ATLAS:2020zzb} overlaid with the simulated best-fit signal counts from the \fermiacc{}'s proposal \texttt{f597e791} for an octet scalar with EFT couplings.}
    \label{fig:proposalf597e791}
\end{figure}

This proposal involves a 1.3 TeV $SU(3)_c$ octet scalar $S^a$ with additional EFT couplings to first-generation Standard Model quarks, arising in the electroweak symmetric phase from
\begin{equation}
\mathcal{L} \supset k_u S^a (\bar{Q}_1 T^a u^c_1) \tilde H_i + k_d S^a (\bar{Q}_1 T^a d_1^c) H_i + {\rm h.c.}
\end{equation}
The scalar mass is chosen to reproduce the observed excess, while the values of the EFT couplings are chosen to yield prompt decays into dijets with a modest width. The interaction structure is chosen to allow production of $S^a$ in association with a $W$ boson; the $S^a$ decays to dijets while the $W$ decays give the lepton required by the signal region. While this model would presumably be subject to complementary constraints from $S^a$ pair production, the proposal is compatible with the information provided in the paper.
The \fermiacc{} then pushed this proposal through the deterministic simulation pipeline to produce the histogram shown in Fig.~\ref{fig:proposalf597e791}.

Taken together, these examples demonstrate the potential of the \fermiacc{} to propose and validate novel hypotheses to explain features in experimental data. Although some of the examples presented above fall short of completely explaining the observed features (typically by failing to reproduce the necessary rate with the chosen model parameters), these single-shot runs of the \fermiacc{} illustrate both its first-pass performance and underline the potential value of iterating the \fermiacc{} over model runs.

\section{\label{sec:conc} Conclusion}

This work details the philosophy, architecture, and performance of the \fermiacc{}, a scaffolded reasoning model built on OpenAI agents designed to autonomously generate and quantitatively validate plausible hypotheses for high energy physics data. In its initial incarnation, the \fermiacc{} produces and tests novel explanations for observed excesses in LHC data. It leverages the rapid progress in general-purpose models while guarding against the shortcomings of modern LLMs by coupling them to deterministic simulation tools. We demonstrate its performance on a range of mild statistical fluctuations in ATLAS and CMS data, including both the historical 750 GeV excess and more recent analyses in a variety of final states.

A central contribution of the \fermiacc{} is to move language models beyond their usual role as chat-based assistants. Instead of generating unconstrained text, the model produces structured outputs that can be passed directly to simulation tools. This turns what is typically an interactive, ad hoc process into one that is reproducible and can be checked step by step, where proposed explanations are not only suggested but also worked out and tested quantitatively. In practice, this makes it possible to explore many competing explanations quickly and in parallel.

There are numerous opportunities for further development. The examples presented in this paper come from single-shot runs of the \fermiacc{}, and in some cases already illustrate the value of iterating over multiple runs. At the level of LHC reinterpretation, the \fermiacc{} would benefit from an analysis validation agent capable of checking its own simulation pipeline and reweighting its fast detector simulation to better emulate the analysis of interest. Given a plausible model explanation for an observed excess, the same architecture can be extended to test relevant limits in related analyses. Whenever model signatures extend beyond the LHC, additional simulation pipelines can be incorporated to evaluate signals in other experiments. Ultimately, some of the challenges associated with recasting and correlating analyses could be alleviated by deploying the \fermiacc{} directly within experimental collaborations.

More broadly, the \fermiacc{} provides a template for verifiable hypothesis generation in high energy physics. Similar adversarial agent configurations may prove useful in the development of AI Einstein and AI Fermi alike, with the potential to significantly accelerate progress in our understanding of Nature.

\begin{acknowledgments}
We would like to thank Zvi Bern, Wahid Bhimji, Kyle Cranmer, Lance Dixon, Simon Knapen, Tongyan Lin, Siddharth Mishra-Sharma, Moritz Muenchmeyer, and David Shih for useful conversations, and Jessica Howard for both essential conversations and collaboration on related work. We are particularly grateful to Kevin Weil and the OpenAI for Science team for support. Parts of this work were performed at the Kavli Institute for Theoretical Physics, supported by the National Science Foundation under Grant No. NSF PHY-1748958. AM was supported by grant GBMF7392 from the Gordon and Betty Moore Foundation. Use was made of computational facilities purchased with funds from the National Science Foundation (CNS-1725797) and administered by the Center for Scientific Computing (CSC). The CSC is supported by the California NanoSystems Institute and the Materials Research Science and Engineering Center (MRSEC; NSF DMR 2308708) at UC Santa Barbara.
\end{acknowledgments}

\bibliographystyle{JHEP}
\bibliography{fermiaccbib}

\appendix

\newpage

\section{Implementation Details of the Model Builder}
\label{app:hypothesis_impl}

This appendix collects the implementation-level details of the hypothesis-generation stage that are omitted from the main text for clarity. In particular, it specifies the structured hypothesis representation used internally by the pipeline, the iterative propose--critique--patch procedure used to refine candidate models, the canonicalized retrieval scheme used to compare new proposals against previously generated hypotheses, and the executable benchmark contract imposed by the downstream UFO-generation stage. These details are included to make the method reproducible and to distinguish the physics content of a generated hypothesis from the software constraints required for automated model construction.

\subsection{Hypothesis Output Schema}

Each hypothesis is returned as a structured object
\[
H = (s, a, F, K, I, C, J, r, U, T, P),
\]
where:
\begin{center}
\begin{tabular}{llp{8.5cm}}
\hline
Symbol & JSON field & Meaning \\
\hline
$s$ & \texttt{analysis\_signature} & One-sentence description of the observable signature, without BSM particle names. \\
$a$ & \texttt{paper\_anchors} & Two to four short evidence anchors from the analysis document. \\
$F$ & \texttt{new\_fields} & List of new BSM fields, including spin, CP label, SM gauge representations, benchmark masses, and self-conjugacy. \\
$K$ & \texttt{new\_kinetic\_terms} & Kinetic and mass terms for the proposed BSM theory content. \\
$I$ & \texttt{new\_interaction\_terms} & Interaction terms needed for the proposed signal mechanism. \\
$C$ & \texttt{new\_couplings} & Benchmark couplings and optional scan ranges. \\
$J$ & \texttt{parameter\_justifications} & Order-of-magnitude justifications for every field mass and coupling. \\
$r$ & \texttt{new\_model\_reasoning} & Brief explanation of why the proposal populates the target signal region(s). \\
$U$ & \texttt{ufo\_contract} & Executable benchmark contract for downstream UFO generation. \\
$T$ & \texttt{bsm\_topology} & Production mode, decay chain, truth objects, reconstructed objects, and missing-energy expectation. \\
$P$ & \texttt{process\_cfg} & \emph{MadGraph} process specification used downstream. \\
\hline
\end{tabular}
\end{center}

A validation rule enforced at generation time is that \texttt{parameter\_justifications} must contain exactly one entry for every element of \texttt{new\_fields} and every element of \texttt{new\_couplings}. This ensures that benchmark values are always accompanied by explicit phenomenological reasoning.

\subsection{Generation Algorithm}

For a given analysis PDF, the hypothesis stage generates $N$ candidate proposals at different sampling temperatures and processes each one independently through an iterative refinement loop:

\begin{enumerate}
\item Generate one candidate hypothesis from the analysis PDF.
\item Retrieve similar prior proposals from the proposal database using canonicalized field-content signatures.
\item Critique the candidate along six axes:
\begin{enumerate}
\item analysis grounding,
\item novelty with respect to the paper and prior proposals,
\item physical consistency,
\item specificity,
\item compatibility with the UFO pipeline,
\item parameter estimation quality.
\end{enumerate}
\item Assign each axis one of \texttt{PASS}, \texttt{REFINE}, or \texttt{FAIL}.
\item Define the overall decision as the maximum-severity label across the six axes.
\item If the overall decision is \texttt{REFINE}, apply the minimal requested patch and return to step 2.
\item Stop when the proposal reaches \texttt{PASS}, reaches \texttt{FAIL}, or exceeds the maximum number of patch rounds.
\end{enumerate}

This procedure yields a set of auditable proposal traces rather than a single opaque model suggestion.

\subsection{Canonicalized Novelty Search}

To compare a new hypothesis against prior generated proposals, each new field is mapped to a canonical signature of the form
\[
\sigma(f) =
(\text{spin class},\,
\text{CP label},\,
\text{color class},\,
\text{electroweak class},\,
Y,\,
\text{self-conjugacy}).
\]

A proposal containing multiple new fields is queried by the multiset of its field signatures. This retrieval step is intentionally coarse: it identifies structurally nearby candidates, after which the agent compares interaction terms, kinetic terms, topology summaries, process definitions, and prior critique outcomes to decide whether the new proposal is genuinely distinct.

\subsection{Executable Benchmark Contract}

The hypothesis representation explicitly separates the \emph{full theory story} from the \emph{executable benchmark}. This distinction is needed because the current downstream model-building path only supports a restricted class of pre-electroweak-symmetry-breaking extensions. In particular, executable benchmarks must avoid:
\begin{itemize}
\item new gauge bosons or gauge sectors,
\item additional vacuum expectation values,
\item explicit broken-phase authoring,
\item custom mixing matrices or mass-diagonalization metadata.
\end{itemize}

The \texttt{ufo\_contract} field records which fields and couplings must survive into the executable benchmark, which couplings may be omitted in a reduced benchmark, and which ultraviolet features are acknowledged but left unsupported.

\subsection{Decision Logging}

Each proposal run is stored as a structured event trace containing:
\begin{itemize}
\item the initial generated hypothesis,
\item each critique object,
\item each patched hypothesis,
\item the final decision and summary.
\end{itemize}

The final proposal summary is inserted into a searchable database together with its field signatures, interaction terms, topology summary, process definition, and final critique outcome. This allows future runs to treat earlier proposals as explicit novelty references.

\section{Implementation Details of Model Construction and Event Generation}
\label{app:ufo_mg5_pipeline}

This appendix gives the implementation-level details of the stage that lies between hypothesis approval and downstream analysis. Unlike the main text, which describes the workflow at a conceptual level, this appendix focuses on the concrete artifacts, validation steps, and control logic used to turn an approved hypothesis into an executable UFO model and then into simulated event samples.

\subsection{Stage Inputs and Handoff Artifacts}

The input to this stage is an approved proposal summary together with its structured benchmark definition. The intermediate stages then write a sequence of proposal-specific artifacts:
\begin{itemize}
\item \texttt{run\_card\_settings.json}: analysis-informed \emph{MadGraph} run-card settings inferred from the PDF;
\item \texttt{param\_map.json}: the map from proposal-level particle and coupling names to executable UFO identifiers, PDG codes, and parameter-card locations;
\item \texttt{ufo\_scope.json}: the reduced executable benchmark contract derived from the proposal;
\item \texttt{build\_status.json}: the status of UFO construction and validation checks, together with warnings and the latest failing diagnostic when relevant;
\item attempt-specific \texttt{BSMExtension.fr.attempt\_\textit{N}.fr} and \texttt{BSMExtension.fr.attempt\_\textit{N}.meta.json} snapshots: the executable extension text and lightweight metadata for each repair attempt;
\item \texttt{ufo\_package.json}: the packaged handoff record for downstream event generation, including the assessed build outcome;

\item \texttt{process\_config.json}: the finalized \emph{MadGraph} process configuration for a benchmark point or scan point;
\item \texttt{parameter\_point.json}: the explicit masses and couplings used in one \emph{MadGraph} run;
\item \texttt{proc.mg5}: the generated \emph{MadGraph} command script.
\end{itemize}

These files make the handoff between stages explicit and allow the workflow to be rerun or audited without recomputing the entire chain.

\subsection{Proposal-Specific Run-Card Inference}

Before the UFO model is built, the pipeline generates a proposal-specific \emph{MadGraph} run card. This helper reads the analysis PDF and a compact textual description of the approved hypothesis and returns a small set of allowed run-card settings, such as:
\begin{itemize}
\item beam type and beam energy;
\item event count;
\item simple object-level cuts, for example photon transverse-momentum or pseudorapidity thresholds;
\item basic separation cuts when these are clearly motivated by the analysis.
\end{itemize}

The run-card helper is intentionally restricted to a short whitelist of robust settings. It does not attempt to tune PDFs or make broad generator-level choices. The goal is not to optimize event generation in a fully general way, but to preserve a reproducible link between the analysis document and the generated benchmark sample.

\subsection{Executable Benchmark Construction}

The approved proposal is next translated into a proposal-specific \emph{FeynRules} extension. This is done by assigning executable identifiers to the new particles and couplings:
\begin{itemize}
\item each new particle is assigned a class identifier and a synthetic PDG code;
\item each new coupling is assigned a parameter-block location, currently in the \texttt{NP} block;
\item proposal-level coupling symbols are mapped to safe \emph{Mathematica/FeynRules} symbols when needed.
\end{itemize}

At the same time, the pipeline derives an executable benchmark contract, stored in \texttt{ufo\_scope.json}. This contract records:
\begin{itemize}
\item the benchmark strategy, i.e.\ whether the executable model is the full proposal or a reduced proxy;
\item the fields that must survive into the benchmark UFO;
\item the couplings that are required for the benchmark process;
\item optional couplings that may be dropped in a reduced benchmark;
\item unsupported features that are acknowledged but left outside the executable model.
\end{itemize}

\subsection{Guardrails Before \emph{FeynRules} Execution}

The UFO-building step is surrounded by a set of guardrails intended to fail fast on common problems before WolframScript is run. These guardrails enforce the restricted contract of the current workflow. In particular, the extension must remain in the unbroken Standard Model gauge basis. The pipeline therefore rejects extensions that contain:
\begin{itemize}
\item broken-phase electroweak fields such as $A$, $Z$, and $W$ written directly in the extension;
\item explicit Higgs vev insertions or electroweak mixing-angle substitutions;
\item new gauge sectors or custom mixing and diagonalization machinery;
\item malformed Higgs bilinears that can leak unphysical \texttt{Phi} or \texttt{Phibar} objects into the exported UFO;
\item fixed-component SU(2) field strengths that are known to cause unphysical \texttt{Wi} objects to leak into the export;
\item kinetic terms that do not use covariant derivatives.
\end{itemize}

These checks are intentionally simple and conservative. They do not prove that the model is physically complete, but they do catch several recurrent failure modes before expensive symbolic processing begins.

\subsection{Iterative UFO Build Procedure}

If the guardrails pass, the pipeline runs the symbolic model-construction stage. The control flow is:
\begin{enumerate}
\item write the proposal-specific \texttt{BSMExtension.fr} file;
\item load the Standard Model base file together with the extension;
\item run \texttt{LoadModel};
\item run \texttt{CheckHermiticity};
\item run \texttt{WriteUFO};
\item sanitize the exported UFO for \emph{MadGraph} use;
\item verify that the benchmark couplings required by the proposal appear in the exported UFO;
\item run a minimal \emph{MadGraph} \texttt{generate} test for the requested hard process.
\end{enumerate}

If a stage fails before the run yields a usable validated UFO, the corresponding diagnostics are passed back into a repair loop and the \emph{FeynRules} extension is rewritten. For several recurrent failure modes, the repair prompt is augmented with targeted hints, for example about duplicate auto-generated mass or width parameters or incorrect orientation of non-self-conjugate fields. The repair loop changes only the executable model text; it does not alter the approved proposal itself. This continues until the UFO build succeeds or the maximum number of repair attempts is reached. Runs that produce a complete UFO but fail a later validation stage are still retained and marked as partial rather than being discarded silently.

\subsection{Validation After UFO Export}

A successful UFO export is not accepted immediately. The exported model is subjected to additional checks:
\begin{itemize}
\item a sanitation pass that removes or rewrites constructs known to break \emph{MadGraph} ingestion;
\item a coupling-survival check, ensuring that the couplings marked as required for the benchmark actually appear in the exported UFO rather than being silently dropped;
\item a minimal \emph{MadGraph} \texttt{generate} test, ensuring that the requested process has at least one valid diagram.
\end{itemize}

The last check is important because a syntactically valid UFO model can still fail to realize the intended benchmark process. For example, a new resonance may exist in the exported model but still lack the tree-level production or decay vertices required by the requested process string. The generate-level validation therefore acts as a final executable guardrail.

\subsection{Translation to \emph{MadGraph} Input}

Once the UFO build succeeds, the pipeline packages the model and constructs the \emph{MadGraph} job. A template process configuration is merged with:
\begin{itemize}
\item the proposal-level process definition;
\item the run-card settings inferred from the analysis PDF;
\item the benchmark parameter values stored in the proposal and mapped through \texttt{param\_map.json}.
\end{itemize}

The executable benchmark parameters are translated into explicit \emph{MadGraph} commands of the form
\[
\texttt{set param\_card BLOCK IDS VALUE},
\]
so that the event-generation step uses the same masses and couplings that were validated during UFO construction.

The resulting \emph{MadGraph} command file has the structure
\begin{enumerate}
\item import the proposal-specific UFO model;
\item define the incoming partons;
\item generate the requested hard process and any additional subprocesses;
\item write the process directory;
\item launch the run;
\item set shower, detector, and analysis switches;
\item apply run-card and parameter-card settings.
\end{enumerate}

\subsection{Event Generation and Optional Detector Simulation}

MadGraph performs the hard-process generation. In the default campaign configuration, the generated parton-level events are then passed to Pythia8 for showering and hadronization. Detector simulation through Delphes is available through the same \emph{MadGraph} steering path, but is currently left off by default in this workflow. The default reason is practical: the main detector-level reinterpretation is performed downstream in MadAnalysis, so this stage is used primarily to produce a validated benchmark event sample rather than a full detector-level analysis product.

\subsection{Benchmark Points and Parameter Scans}

The same infrastructure supports both single-point and multi-point event generation. By default, stage 3 runs only the benchmark point recorded in the proposal. However, proposals may also contain suggested scan ranges for masses and couplings. These suggestions are treated as advisory. If a parameter scan is requested, the pipeline:
\begin{enumerate}
\item reads the suggested scan ranges from the proposal;
\item builds a Cartesian grid of scan points;
\item rescales the per-dimension point counts if necessary to respect a global campaign-level cap on the total number of points;
\item writes a scan-plan file and launches one \emph{MadGraph} run per retained point.
\end{enumerate}

This cap is an implementation guardrail intended to prevent the parameter scan from growing without bound. The benchmark point remains the default reference point even when a scan is available.

\subsection{Status Recording}

The outputs of this stage are stored in proposal-specific manifests. In particular, the manifest records:
\begin{itemize}
\item the list of UFO build attempts for a proposal;
\item the latest successful UFO package;
\item the list of \emph{MadGraph} runs performed for that proposal;
\item the latest successful \emph{MadGraph} run;
\item warnings produced by non-fatal stages such as optional Lagrangian rendering.
\end{itemize}

This design makes the workflow rerun-safe and keeps the executable state of each approved proposal explicit.

\section{Agentic Recasting Pipeline}
\label{app:agentic-recasting-pipeline}

The recasting framework is organized as a staged, agentic pipeline that transforms an analysis document into an executable reinterpretation package and corresponding statistical outputs. A central design choice is the use of a structured \texttt{AnalysisTemplate} as the shared intermediate representation. This template stores the analysis metadata, reconstructed objects, signal and auxiliary regions, numerical statistical inputs, and generated backend code. All major stages operate on this shared object, which reduces ambiguity and makes intermediate failures easier to localize.

\subsection{Main Agents}

The main model-driven components of the pipeline are:
\begin{itemize}
    \item \textbf{\texttt{AnalysisPlanner}}, which extracts the semantic structure of the analysis from the source document, including the event topology, region definitions, and limit-setting requirements;
    \item \textbf{\texttt{DataSourceAgent}}, which identifies external numerical sources needed for recasting;
    \item \textbf{\texttt{AnalysisCppGenerator}}, which emits executable backend code, currently centered on MadAnalysis~5.
\end{itemize}

These agents are complemented by deterministic mapping, validation, archival, execution, and statistics stages.

\subsection{Data-Source Layer}

The \texttt{DataSourceAgent} is responsible for source discovery rather than final numerical arbitration. In practice, it searches over several source classes:
\begin{itemize}
    \item HEPData records and other structured public datasets,
    \item official analysis pages and collaboration-hosted plots,
    \item digitized spectra and plot-derived numerical series,
    \item user-provided mappings such as curated CSV or JSON tables.
\end{itemize}

This stage combines language-model reasoning with explicit external tools. Web search is used to locate HEPData records, official pages, and relevant figures. When structured public tables exist, a dedicated HEPData reader is used to retrieve them. When the required information is only available graphically, a digitization component is used to extract approximate numerical series from spectra, fit curves, or uncertainty bands. User-provided mappings are treated as first-class inputs and can override weaker public-source matches when needed.

The output of this stage is a structured source plan, not yet the final signal-region data.

\subsection{Deterministic Resolution and Code Generation}

After source discovery, deterministic mapping stages align the candidate inputs to the signal-region structure defined in the analysis template. A subsequent resolution stage determines which observed values, background expectations, and background uncertainties populate the canonical numerical fields used for recasting.

The \texttt{AnalysisCppGenerator} then converts the structured analysis into executable backend code. In the current implementation this primarily targets \emph{MadAnalysis~5}. The generated package includes the analyzer C++ implementation, region and histogram declarations, backend metadata, and archived run artifacts. A deterministic backend-preparation layer then performs topology normalization, metadata writing, and consistency validation between the generated code and the region/statistics definitions.

\subsection{Execution and Archival}

The generated backend package is archived in a run-specific directory, mirrored into a live \emph{MadAnalysis~5} installation, compiled, and run on \emph{Delphes}-level event samples. Native \emph{MadAnalysis~5} statistics are then used to compute efficiencies, best-region summaries, and CL$_s$-related outputs. The full run is persisted together with tool outputs, generated code, logs, and statistical artifacts.

This archival structure is an important part of the architecture: it makes each stage inspectable and supports targeted debugging when a later stage fails despite earlier stages succeeding.

\subsection{Architectural Perspective}

The workflow is agentic in the practical sense that each major stage has a specialized role and produces an explicit artifact. This allows failures to be localized to planning, data acquisition, statistical-input resolution, code generation, or runtime execution, rather than being hidden in a single opaque generation step.

At the same time, experience with the current implementation shows that region topology, cut naming, and backend metadata must ultimately be owned by a deterministic core. The long-term architectural direction is therefore to combine an agentic front end for planning and source discovery with a stronger deterministic backbone for topology construction, backend emission, validation, and staged reruns.

\section{Dissecting the Reasoning Model}
In this appendix we show snippets of the reasoning model outputs in order to provide insight into its internal workings. We use the proposal {\texttt{544c3421}} for the CMS Diphoton excess as an example. The specific reasoning steps are a result of internal processing, following from a general prompt. We will see that the model already knows a remarkable amount of effective field theory.
It performs first estimates and checks for logical consistency and obvious constraints.

\subsection{New fields and couplings}
The model generates a proposal with new fields, their interactions and couplings, along with an initial estimate of mass and coupling values, and recommended scan ranges. This parameter estimation is justified in detail (see below for examples). These initial values are already reasonable guesses for explaining the signal, and are further refined by passing on to the deterministic pipeline.
\jsonblock{
\jk{analysis\_signature}: \jv{"A narrow high-mass diphoton resonance reconstructed from two isolated energetic photons, split into barrel-barrel and barrel-endcap categories, with no required missing transverse momentum and otherwise inclusive in additional activity."}
}

\jsonblock{
\jk{new\_fields}: [\{\\
\hspace*{1.5em}\jk{name}: \jv{"A"},\\
\hspace*{1.5em}\jk{spin}: \jv{"0"},\\
\hspace*{1.5em}\jk{cp}: \jv{"odd"},\\
\hspace*{1.5em}\jk{sm\_reps}: \jv{\{"su3": "1", "su2": "1", "u1y": 0.0\}},\\
\hspace*{1.5em}\jk{mass\_GeV}: \jv{750.0},\\
\hspace*{1.5em}\jk{mass\_scan}: \jv{\{"min\_value": 500.0, "max\_value": 2000.0, "spacing": "linear", "suggested\_points": 7\}},\\
\hspace*{1.5em}\jk{self\_conjugate}: \jv{true}\\
\}, \{\\
\hspace*{1.5em}\jk{name}: \jv{"T"},\\
\hspace*{1.5em}\jk{spin}: \jv{"1/2"},\\
\hspace*{1.5em}\jk{cp}: \jv{"not\_applicable"},\\
\hspace*{1.5em}\jk{sm\_reps}: \jv{\{"su3": "3", "su2": "1", "u1y": 0.6666666667\}},\\
\hspace*{1.5em}\jk{mass\_GeV}: \jv{1200.0},\\
\hspace*{1.5em}\jk{mass\_scan}: \jv{\{"min\_value": 900.0, "max\_value": 2000.0, "spacing": "linear", "suggested\_points": 6\}},\\
\hspace*{1.5em}\jk{self\_conjugate}: \jv{false}\\
\}]
}

\jsonblock{
\jk{new\_couplings}: [\{\\
\jline{1.5em}{\jk{symbol}: \jv{yT}, \jk{value}: \jv{10.0}, \jk{mass\_dimension}: \jv{0},}
\jline{1.5em}{\jk{scan}: \jv{\{min\_value: 4.0, max\_value: 12.0, spacing: linear, suggested\_points: 5\}}}
\}, \{\\
\jline{1.5em}{\jk{symbol}: \jv{yH}, \jk{value}: \jv{0.1}, \jk{mass\_dimension}: \jv{0},}
\jline{1.5em}{\jk{scan}: \jv{\{min\_value: 0.03, max\_value: 0.3, spacing: linear, suggested\_points: 5\}}}
\}, \{\\
\jline{1.5em}{\jk{symbol}: \jv{cAG}, \jk{value}: \jv{3e-05}, \jk{mass\_dimension}: \jv{-1},}
\jline{1.5em}{\jk{scan}: \jv{\{min\_value: 1e-05, max\_value: 6e-05, spacing: linear, suggested\_points: 6\}}}
\}, \{\\
\jline{1.5em}{\jk{symbol}: \jv{cAB}, \jk{value}: \jv{9e-06}, \jk{mass\_dimension}: \jv{-1},}
\jline{1.5em}{\jk{scan}: \jv{\{min\_value: 3e-06, max\_value: 2e-05, spacing: linear, suggested\_points: 6\}}}
\}]
}

\subsection{Parameter justifications}
The model produces detailed justifications for choices of parameters it suggests to use down the pipeline. For the masses it scans the paper for relevant details, but also uses internal model consistency to ensure that the dominant decay of the field $A$ is at loop-level. It also follows the prompt to not simply return the model interpretation included in the analysis.
\jsonblock{
\jk{parameter\_justifications}: [\{\\
\jline{1.5em}{\jk{target\_kind}: \jv{field\_mass}, \jk{target\_name}: \jv{A}, \jk{oom\_value}: \jv{3},}
\jline{1.5em}{\jk{calculation\_steps}: [}
\jline{3em}{\jv{analysis inspects the diphoton mass spectrum well above 500 GeV},}
\jline{3em}{\jv{the paper notes the largest early-data fluctuation near 760 GeV for a narrow resonance hypothesis},}
\jline{3em}{\jv{choose a representative in-window benchmark mA = 750 GeV, i.e. O(10\string^3 GeV), while keeping the model distinct from the paper's spin-2 interpretation}}
\jline{1.5em}{]}
\}, \{\\
\jline{1.5em}{\jk{target\_kind}: \jv{field\_mass}, \jk{target\_name}: \jv{T}, \jk{oom\_value}: \jv{3},}
\jline{1.5em}{\jk{calculation\_steps}: [}
\jline{3em}{\jv{to keep A -> T Tbar closed, require mT > mA/2 = 375 GeV},}
\jline{3em}{\jv{a heavier colored fermion also justifies matching to dimension-5 Agg and ABB operators},}
\jline{3em}{\jv{choose mT = 1.2e3 GeV so the EFT picture is reasonable while yT can still generate useful effective couplings without exceeding 4pi}}
\jline{1.5em}{]}
\},
}
\vspace{10pt}

The model estimates the effective coupling required for fitting the signal, performs a naive loop estimate for the UV coupling needed to generate the EFT coupling required, and worries about  staying away from strong coupling.
\jsonblock{
\{\\
\jline{1.5em}{\jk{target\_kind}: \jv{coupling}, \jk{target\_name}: \jv{yT}, \jk{oom\_value}: \jv{1},}
\jline{1.5em}{\jk{calculation\_steps}: [}
\jline{3em}{\jv{target collider-level production needs cAG of a few x 10\string^-5 GeV\string^-1},}
\jline{3em}{\jv{heavy-fermion matching gives cAG $\sim$ alpha\_s yT /(8 pi mT)},}
\jline{3em}{\jv{with alpha\_s $\sim$ 0.09, mT = 1.2e3 GeV, and cAG = 3e-5 GeV\string^-1, solve yT $\sim$ cAG (8 pi mT)/alpha\_s $\sim$ 10},}
\jline{3em}{\jv{10 is strong but still below 4pi, so use yT = 10}}
\jline{1.5em}{]}
\},}

\vspace{10pt}

The model uses its estimates for the couplings in the EFT to estimate the production cross section and decay widths.
\jsonblock{
\{\\
\jline{1.5em}{\jk{target\_kind}: \jv{coupling}, \jk{target\_name}: \jv{cAG}, \jk{oom\_value}: \jv{-5},}
\jline{1.5em}{\jk{calculation\_steps}: [}
\jline{3em}{\jv{use heavy-T matching cAG $\sim$ alpha\_s yT /(8 pi mT)},}
\jline{3em}{\jv{alpha\_s $\sim$ 0.09, yT = 10, mT = 1.2e3 GeV gives cAG $\sim$ 0.09*10/(8*pi*1200) $\sim$ 3e-5 GeV\string^-1},}
\jline{3em}{\jv{for L $\sim$ (cAG/4) A G Gtilde, estimate Gamma(A -> gg) $\sim$ cAG\string^2 mA\string^3 /(8 pi) $\sim$ 1e-2 GeV at mA = 750 GeV},}
\jline{3em}{\jv{this gives Gamma/mA $\sim$ 2e-5, so the resonance remains narrow while gg production can be O(10\string^2 fb) at 13 TeV}}
\jline{1.5em}{]}
\},
}
\jsonblock{
\{\\
\jline{1.5em}{\jk{target\_kind}: \jv{coupling}, \jk{target\_name}: \jv{cAB}, \jk{oom\_value}: \jv{-5},}
\jline{1.5em}{\jk{calculation\_steps}: [}
\jline{3em}{\jv{for an SU(2)\_L-singlet T with Y = 2/3, estimate cAB $\sim$ alpha\_Y Nc Y\string^2 yT /(4 pi mT)},}
\jline{3em}{\jv{taking alpha\_Y $\sim$ 0.01, Nc Y\string^2 = 3*(4/9), yT = 10, and mT = 1.2e3 GeV gives cAB $\sim$ 9e-6 GeV\string^-1},}
\jline{3em}{\jv{after EWSB, cAaa $\sim$ cAB cos\string^2(thetaW) $\sim$ 7e-6 GeV\string^-1, so Gamma(A -> aa) $\sim$ cAaa\string^2 mA\string^3 /(64 pi) $\sim$ 1e-4 GeV},}
\jline{3em}{\jv{including the related Zgamma and ZZ modes from the same B\_\{mu nu\} operator, the electroweak widths stay well below Gamma(A -> gg), giving BR(A -> aa) of order 10\string^-2 and sigma(pp -> A -> aa) of order 1 fb, enough to populate the search window while staying narrow}}
\jline{1.5em}{]}
\}]
}

\subsection{Model interpretation and passing on to collider tools}
The reasoning notes that there is an existing EFT model in its internal database that this theory matches on to, and that the novelty of this proposal lies in providing a UV completion for that theory.

\jsonblock{
\jk{new\_model\_reasoning}: \jv{"A CP-odd singlet produced by gluon fusion and decaying to two photons gives exactly the narrow s-channel diphoton bump targeted by the analysis. The collider-facing content is intentionally not new: it reuses the existing CP-odd singlet gg -> A -> aa EFT benchmark. The only novel layer in this submission is the UV completion, where a heavy vectorlike quark T with couplings yT and yH generates the effective A G Gtilde and A B Btilde operators after matching. Therefore this entry should be stored as a UV-completion variant linked to that existing EFT benchmark, not as a distinct standalone collider benchmark."}
}

\vspace{10pt}

The model then produces a UFO contract for the reduced collider benchmark that is passed into the deterministic pipeline. It retains only the effective couplings and the ``light'' particle relevant for the phenomenology, while explicitly leaving out UV-completion features such as the $T$-quark and its mixing with the top quark, which are not needed for the target signature. It also notes that explicit one-loop matching should be added in the future to obtain a more robust mapping between the UV theory and the EFT.

\jsonblock{
\jk{ufo\_contract}: \{\\
\jline{1.5em}{\jk{benchmark\_strategy}: \jv{minimal\_benchmark},}
\jline{1.5em}{\jk{required\_fields}: \jv{[A]},}
\jline{1.5em}{\jk{required\_couplings}: \jv{[cAG, cAB]},}
\jline{1.5em}{\jk{optional\_couplings}: \jv{[yT, yH]},}
\jline{1.5em}{\jk{unsupported\_features}: [}
\jline{3em}{\jv{explicit one-loop matching from the heavy vectorlike quark T onto cAG and cAB within the pre-EWSB tree-level UFO pipeline},}
\jline{3em}{\jv{the UV-completion field T and the UV parameters yT and yH are not part of the reduced collider UFO and should remain outside the benchmark build},}
\jline{3em}{\jv{EWSB-induced top-T mixing and the resulting mass-basis decay structure of T are not needed for process\_cfg and may be omitted from the benchmark UFO},}
\jline{3em}{\jv{the UV loop completion and the reduced effective operators should not be simulated simultaneously in one benchmark build},}
\jline{3em}{\jv{manual addition of separate mass-basis Aaa, Aaz, or Azz operators on top of the gauge-basis ABB operator is outside the intended benchmark and would double count the standard electroweak rotation},}
\jline{3em}{\jv{the benchmark-level collider content coincides with the existing CP-odd singlet diphoton EFT entry, so this hypothesis must be treated only as that benchmark's UV-completion variant rather than as a new standalone collider benchmark}}
\jline{1.5em}{],}
\jline{1.5em}{\jk{notes}: \jv{Build only the reduced CP-odd singlet EFT proxy with A, cAG, and cAB, and link it explicitly to the existing gg -> A -> aa benchmark; the novelty here is solely the heavy-T UV completion.}}
\}
}
\vspace{10pt}
The summary notes whether this proposal is to be accepted or resubmitted to the agent for refinement based on the outputs and the critique.
\jsonblock{
\jk{critique\_summary}: \jv{"The proposal is well grounded in the CMS high-mass diphoton search and is physically/pipeline consistent as a reduced CP-odd singlet EFT benchmark. Its collider-facing benchmark matches an existing EFT entry, but the added vectorlike-quark UV completion is a meaningful full-story distinction and is already correctly framed as a linked UV-completion variant rather than a new standalone benchmark."},\\
\jk{\_\_meta\_\_}: \{\\
\hspace*{1.5em}\jk{id}: \jv{"544c3421-c756-498c-a9b9-22b37d755c21"},\\
\hspace*{1.5em}\jk{decision}: \jv{"PASS"},\\
\hspace*{1.5em}\jk{temperature}: \jv{0.7}\\
\}
}

\end{document}